\documentclass[11pt,a4paper]{article}
\usepackage{amsfonts}
\usepackage{amsmath,amssymb}
\usepackage{epsfig,graphicx}

\input{tcilatex}

\begin{document}

\begin{center}
{\LARGE \ }{\huge On emergent SUSY gauge theories}

\bigskip \bigskip

\textbf{J.L.~Chkareuli}$^{1,2}$

$^{1}$\textit{Center for Elementary Particle Physics, Ilia State University,
0162 Tbilisi, Georgia\ \vspace{0pt}\\[0pt]
}

$^{2}$\textit{E. Andronikashvili} \textit{Institute of Physics, 0177
Tbilisi, Georgia\ }

\bigskip

\bigskip

\bigskip

\bigskip

\bigskip

\bigskip

\bigskip

\textbf{Abstract}
\end{center}

We present the basic features of emergent SUSY gauge theories where an
emergence of gauge bosons as massless vector Nambu-Goldstone modes is
triggered by the spontaneously broken supersymmetry rather than the
physically manifested Lorentz violation. We start considering the
supersymmetric QED model extended by an arbitrary polynomial potential of
massive vector superfield that induces the spontaneous SUSY violation in the
visible sector. As a consequence, a massless photon appears as a companion
of a massless photino emerging as a goldstino in the tree approximation, and
remains massless due to the simultaneously generated special gauge
invariance. This invariance is only restricted by the supplemented vector
field constraint invariant under supergauge transformations. Meanwhile,
photino being mixed with another goldstino appearing from a spontaneous SUSY
violation in the hidden sector largely turns into the light
pseudo-goldstino. Such pseudo-goldstonic photinos considered in an extended
supersymmetric Standard Model framework are of a special observational
interest that, apart from some indication of the QED emergence nature, may
appreciably extend the scope of SUSY breaking physics being actively studied
in recent years. \bigskip

\bigskip

\bigskip

\bigskip

\bigskip

\bigskip

\bigskip

\bigskip

\bigskip

\bigskip

\bigskip

\bigskip

\bigskip

\begin{center}
{\tiny Invited talk at the International Workshop "What Comes Beyond the
Standard Model?" (14-21 July 2013, Bled, Slovenia)}
\end{center}

\thispagestyle{empty}\newpage

\section{Introduction}

It is long believed that spontaneous Lorentz invariance violation (SLIV) may
lead to an emergence of massless Nambu-Goldstone modes \cite{NJL} which are
identified with photons and other gauge fields appearing in the Standard
Model. This idea \cite{bjorken} supported by a close analogy with the
dynamical origin of massless particle excitations for spontaneously broken
internal symmetries has gained new impetus \cite{cfn,jb,kraus,jen,bluhm} in
recent years.

In this connection, one important thing to notice is that, in contrast to
the spontaneous violation of internal symmetries, SLIV seems not to
necessarily imply a physical breakdown of Lorentz invariance. Rather, when
appearing in a gauge theory framework, this may ultimately result in a
noncovariant gauge choice in an otherwise gauge invariant and Lorentz
invariant theory. In substance, the SLIV ansatz, due to which the vector
field develops a vacuum expectation value (vev)
\begin{equation}
<A_{\mu }(x)>\text{ }=n_{\mu }M  \label{vev1}
\end{equation}%
(where $n_{\mu }$ is a properly-oriented unit Lorentz vector, $n^{2}=n_{\mu
}n^{\mu }=\pm 1$, while $M$ is the proposed SLIV scale), may itself be
treated as a pure gauge transformation with a gauge function linear in
coordinates, $\omega (x)=$ $n_{\mu }x^{\mu }M$. From this viewpoint gauge
invariance in QED leads to the conversion of SLIV into gauge degrees of
freedom of the massless Goldstonic photon emerged.

A good example for such a kind of the "inactive" SLIV is provided by the
nonlinearly realized Lorentz symmetry for underlying vector field $A_{\mu
}(x)$ through the length-fixing constraint%
\begin{equation}
A_{\mu }A^{\mu }=n^{2}M^{2}\text{ .}  \label{const}
\end{equation}%
This constraint in the gauge invariant QED framework was first studied by
Nambu a long ago \cite{nambu}, and in more detail in recent years \cite%
{az,kep,jej,urr,gra}. The constraint (\ref{const}) is in fact very similar
to the constraint appearing in the nonlinear $\sigma $-model for pions \cite%
{GL}, $\sigma ^{2}+\pi ^{2}=f_{\pi }^{2}$, where $f_{\pi }$ is the pion
decay constant. Rather than impose by postulate, the constraint (\ref{const}%
) may be implemented into the standard QED Lagrangian $L_{QED}$ through the
invariant Lagrange multiplier term

\begin{equation}
L_{tot}=L_{QED}-\frac{\lambda }{2}\left( A_{\mu }A^{\mu }-n^{2}M^{2}\right)
\label{lag}
\end{equation}%
provided that initial values for all fields (and their momenta) involved are
chosen so as to restrict the phase space to values with a vanishing
multiplier function $\lambda (x)$, $\lambda =0$ \footnote{%
Otherwise, as was shown in \cite{vru} (see also \cite{urr}), it might be
problematic to have the ghost-free QED model with a positive Hamiltonian.}.

One way or another, the constraint (\ref{const}) means in essence that the
vector field $A_{\mu }$ develops the vev (\ref{vev1}) and Lorentz symmetry $%
SO(1,3)$ breaks down to $SO(3)$ or $SO(1,2)$ depending on whether the unit
vector $n_{\mu }$ is time-like ($n^{2}>0$) or space-like ($n^{2}<0$). The
point, however, is that, in sharp contrast to the nonlinear $\sigma $ model
for pions, the nonlinear QED theory, due to gauge invariance in the starting
Lagrangian $L_{QED}$, ensures that all the physical Lorentz violating
effects turn out to be non-observable. Actually, as was shown in the tree
\cite{nambu} and one-loop approximations \cite{az}, the nonlinear constraint
(\ref{const}) implemented as a supplementary condition appears in essence as
a possible gauge choice for the vector field $A_{\mu }$, while the $S$%
-matrix remains unaltered under such a gauge convention. So, as generally
expected, the inactive SLIV inspired by the length-fixing constraint (\ref%
{const}), while producing an ordinary photon as a true Goldstonic vector
boson ($a_{\mu }$)
\begin{equation}
A_{\mu }=a_{\mu }+n_{\mu }(M^{2}-n^{2}a^{2})^{\frac{1}{2}}\text{ , \ }n_{\mu
}a_{\mu }=0\text{ \ \ (}a^{2}\equiv a_{\mu }a^{\mu }\text{) ,}  \label{gol}
\end{equation}%
leaves physical Lorentz invariance intact\footnote{%
Indeed, the nonlinear QED contains a plethora of Lorentz and $CPT$ violating
couplings when it is expressed in terms of the pure Goldstonic photon modes $%
a_{\mu }$. However, the contributions of all these couplings to physical
processes completely cancel out among themselves.}. Later similar result was
also confirmed for spontaneously broken massive QED \cite{kep}, non-Abelian
theories \cite{jej} and tensor field gravity \cite{gra}.

From this point of view, emergent gauge theories induced by the inactive
SLIV mechanism are in fact indistinguishable from conventional gauge
theories. Their Goldstonic nature could only be seen when taking the gauge
condition of the length-fixing constraint type (\ref{const}). Any other
gauge, e.g. Coulomb gauge, is not in line with Goldstonic picture, since it
breaks Lorentz invariance in an explicit rather than spontaneous way. As to
an observational evidence in favor of emergent theories the only way for
inactive SLIV to cause physical Lorentz violation would be if gauge
invariance in these theories appeared slightly broken in an explicit, rather
than spontaneous, way. Actually, such a gauge symmetry breaking, induced by
some high-order operators, leads in the presence of SLIV to deformed
dispersion relations for matter and gauge fields involved. This effect
typically appears proportional to powers of the ratio $M/M_{P}$, so that for
some high value of the SLIV\ scale $M$ it may become physically observable
even at low energies. Though one could speculate about some generically
broken or partial gauge symmetry \cite{par}, this seems to be too high price
for an actual Lorentz violation which \ may stem from SLIV\footnote{%
In this connection, the simplest possibility could be a conventional QED
Lagrangian extended by the vector field potential energy terms, $L=L_{QED}-%
\frac{\boldsymbol{\lambda }}{4}\left( A_{\mu }A^{\mu }-n^{2}M^{2}\right) ^{2}
$, where $\boldsymbol{\lambda }$ is a coupling constant. This Lagrangian
being sometimes referred to as the \textquotedblleft
bumblebee\textquotedblright\ model (see \cite{bluhm} and references therein)
is in a sense a linear version of the nonlinear QED appearing in the limit $%
\boldsymbol{\lambda }\rightarrow \infty $. Actually, both of models are
physically equivalent in the infrared energy domain, where the Higgs mode is
considered infinitely massive. However, as we see shortly, whereas the
nonlinear QED model successfully matches supersymmetry, the
\textquotedblleft bumblebee\textquotedblright\ model cannot be conceptually
realized in the SUSY context.}. And, what is more, is there really any
strong theoretical reason left for Lorentz invariance to be physically
broken, if the Goldstonic gauge fields are anyway generated through the
\textquotedblleft safe\textquotedblright\ inactive SLIV models which recover
conventional Lorentz invariance?

Nevertheless, it may turn out that SLIV is not the only reason why massless
photons could dynamically appear, if spacetime symmetry is further enlarged.
In this connection, special interest may be related to supersymmetry.
Actually, as we try to show below, the situation is changed remarkably in
the SUSY inspired emergent models which, in contrast to non-SUSY analogues,
could naturally have some clear observational evidence. We argue that a
generic source for massless photons may be spontaneously broken
supersymmetry rather than physically manifested spontaneous Lorentz
violation \cite{jc}. Towards this end, we consider supersymmetric QED model
extended by an arbitrary polynomial potential of massive vector superfield
that induces the spontaneous SUSY violation\footnote{%
It is worth noting that all the basic arguments related to the present QED
example can be then straightforwardly extended to the Standard Model.}. As a
consequence, a massless photon emerges as a companion of a massless photino
being Goldstone fermion in the broken SUSY phase in the visible sector
(section 2). Remarkably, this masslessness appearing at the tree level is
further protected against radiative corrections by the simultaneously
generated special gauge invariance. This invariance is only restricted by
the supplemented vector field constraint invariant under supergauge
transformations (section 3). Meanwhile, photino being mixed with another
goldstino appearing from a spontaneous SUSY violation in the hidden sector
largely turns into the light pseudo-goldstino whose physics seems to be of
special interest (section 4). And finally, we conclude (section 5).

\section{Extended supersymmetric QED}

We now consider the supersymmetric QED extended by an arbitrary polynomial
potential of \ a general vector superfield $V(x,\theta ,\overline{\theta })$
which in the standard parametrization \cite{wess} has a form
\begin{eqnarray}
V(x,\theta ,\overline{\theta }) &=&C(x)+i\theta \chi -i\overline{\theta }%
\overline{\chi }+\frac{i}{2}\theta \theta S-\frac{i}{2}\overline{\theta }%
\overline{\theta }S^{\ast }  \notag \\
&&-\theta \sigma ^{\mu }\overline{\theta }A_{\mu }+i\theta \theta \overline{%
\theta }\overline{\lambda ^{\prime }}-i\overline{\theta }\overline{\theta }%
\theta \lambda ^{\prime }+\frac{1}{2}\theta \theta \overline{\theta }%
\overline{\theta }D^{\prime },  \label{par}
\end{eqnarray}%
where its vector field component $A_{\mu }$ is usually associated with a
photon. Note that, apart from the conventional photino field $\lambda $ and
the auxiliary $D$ field , the superfield (\ref{par}) contains in general the
additional degrees of freedom in terms of the dynamical $C$ and $\chi $
fields and nondynamical complex scalar field $S$ (we have used the brief
notations, $\lambda ^{\prime }=\lambda +\frac{i}{2}\sigma ^{\mu }\partial
_{\mu }\overline{\chi }$ \ and $D^{\prime }=D+\frac{1}{2}\partial ^{2}C$
with $\sigma ^{\mu }=(1,\overrightarrow{\sigma })$ and $\overline{\sigma }%
^{\mu }=(1,-\overrightarrow{\sigma })$). The corresponding SUSY invariant
Lagrangian may be written as%
\begin{equation}
\mathcal{L}=L_{SQED}+\sum_{n=1}b_{n}V^{n}|_{D}  \label{slag}
\end{equation}%
where terms in this sum ($b_{n}$ are some constants) for the vector
superfield (\ref{par}) are given through the $V^{n}|_{D}$ expansions into
the component fields . It can readily be checked that the first term in this
expansion appears to be the known Fayet-Iliopoulos $D$-term, while other
terms only contain bilinear, trilinear and quadrilinear combination of the
superfield components $A_{\mu }$, $S$, $\lambda $ and $\chi $, respectively%
\footnote{%
Note that all terms in the sum in (\ref{slag}) except Fayet-Iliopoulos $D$%
-term\ explicitly break gauge invariance which is then recovered for
Goldstonic gauge modes. Without loss of generality, we may restrict
ourselves to the third degree superfield polynomial in the Lagrangian $%
\mathcal{L}$ (\ref{slag}) to eventually have a theory with dimesionless
coupling constants for component fields. However, for completeness sake, it
seems better to proceed with a general case.}. Actually, there appear
higher-degree terms for the scalar field component $C(x)$ only. Expressing
them all in terms of the $C$ field polynomial%
\begin{equation}
P(C)=\sum_{n=1}\frac{n}{2}b_{n}C^{n-1}(x)  \label{pot}
\end{equation}%
and its first three derivatives
\begin{equation}
P_{C}^{\prime }\equiv \frac{\partial P}{\partial C}\text{ , \ \ }%
P_{C}^{\prime \prime }\equiv \frac{\partial ^{2}P}{\partial C^{2}}\text{ , \
\ }P_{C}^{\prime \prime \prime }\equiv \frac{\partial ^{3}P}{\partial C^{3}}%
\text{ }  \label{dd}
\end{equation}%
one has for the whole Lagrangian $\mathcal{L}$
\begin{eqnarray}
\mathcal{L} &=&-\text{ }\frac{1}{4}F^{\mu \nu }F_{\mu \nu }+i\lambda \sigma
^{\mu }\partial _{\mu }\overline{\lambda }+\frac{1}{2}D^{2}  \notag \\
&&+\text{ }P\left( D+\frac{1}{2}\partial ^{2}C\right) +P_{C}^{\prime }\left(
\frac{1}{2}SS^{\ast }-\chi \lambda ^{\prime }-\overline{\chi }\overline{%
\lambda ^{\prime }}-\frac{1}{2}A_{\mu }A^{\mu }\right)  \notag \\
&&+\text{ }\frac{1}{2}P_{C}^{\prime \prime }\left( \frac{i}{2}\overline{\chi
}\overline{\chi }S-\frac{i}{2}\chi \chi S^{\ast }-\chi \sigma ^{\mu }%
\overline{\chi }A_{\mu }\right) +\frac{1}{8}P_{C}^{\prime \prime \prime
}(\chi \chi \overline{\chi }\overline{\chi })\text{ .}  \label{lag3}
\end{eqnarray}%
where, for more clarity, we still omitted matter superfields in the model
reserving them for section 4. As one can see, extra degrees of freedom
related to the $C$ and $\chi $ component fields in a general vector
superfield $V(x,\theta ,\overline{\theta })$ appear through the potential
terms in (\ref{lag3}) rather than from the properly constructed
supersymmetric field strengths, as is appeared for the vector field $A_{\mu
} $ and its gaugino companion $\lambda $.

Varying the Lagrangian $\mathcal{L}$ with respect to the $D$ field we come
to
\begin{equation}
D=-P(C)  \label{d}
\end{equation}%
that finally gives the following potential energy for the field system
considered
\begin{equation}
U(C)=\frac{1}{2}[P(C)]^{2}\text{ .}  \label{pot1}
\end{equation}%
The potential (\ref{pot1}) may lead to the spontaneous SUSY breaking in the
visible sector provided that the polynomial $P$ (\ref{pot}) has no real
roots, while its first derivative has,
\begin{equation}
P\neq 0\text{ , \ }P_{C}^{\prime }=0.\text{\ }  \label{der}
\end{equation}%
This requires $P(C)$ to be an even degree polynomial with properly chosen
coefficients $b_{n}$ in (\ref{pot}) that will force its derivative $%
P_{C}^{\prime }$ to have at least one root, $C=C_{0}$, in which the
potential (\ref{pot1}) is minimized and supersymmetry is spontaneously
broken. As an immediate consequence, that one can readily see from the
Lagrangian $\mathcal{L}$ (\ref{lag3}), a massless photino $\lambda $ being
Goldstone fermion in the broken SUSY phase make all the other component
fields in the superfield $V(x,\theta ,\overline{\theta }),$ including the
photon, to also become massless. However, the question then arises whether
this masslessness of photon will be stable against radiative corrections
since gauge invariance is explicitly broken in the Lagrangian (\ref{lag3}).
We show below that it may the case if the vector superfield $V(x,\theta ,%
\overline{\theta })$ would appear to be properly constrained.

\section{Constrained vector superfield}

We have seen above that the vector field $A_{\mu }$ may only appear with
bilinear mass terms in the polynomially extended Lagrangian (\ref{lag3}).
Hence it follows that the \textquotedblleft bumblebee\textquotedblright\
model mentioned above$^{4}$ with nontrivial vector field potential
containing both a bilinear mass term and a quadrilinear stabilizing term can
in no way be realized in the SUSY context. Meanwhile, the nonlinear QED
model, as will become clear below, successfully matches supersymmetry.

Let us constrain our vector superfield $V(x,\theta ,\overline{\theta })$ by
analogy with constrained vector field in the nonlinear QED model (see (\ref%
{lag})). This can be done again through the invariant Lagrange multiplier
term simply adding it to the above Lagrangian (\ref{slag})
\begin{equation}
\mathcal{L}_{tot}=\mathcal{L}+\frac{1}{2}\Lambda (V-C_{0})^{2}|_{D}
\label{ext}
\end{equation}%
where $\Lambda (x,\theta ,\overline{\theta })$ is some auxiliary vector
superfield, while $C_{0}$ is the constant background value of the $C$ field
for which potential $U$ (\ref{pot1}) has the SUSY breaking minimum (\ref{der}%
) in the visible sector.

We further find for the Lagrange multiplier term in (\ref{ext}) that
(denoting $\widetilde{C}\equiv C-C_{0}$)%
\begin{eqnarray}
\Lambda (V-C_{0})^{2}|_{D} &=&{\large C}_{\Lambda }\left[ \widetilde{C}%
D^{\prime }+\left( \frac{1}{2}SS^{\ast }-\chi \lambda ^{\prime }-\overline{%
\chi }\overline{\lambda ^{\prime }}-\frac{1}{2}A_{\mu }A^{\mu }\right) %
\right]  \notag \\
&&+\text{ }{\large \chi }_{\Lambda }\left[ 2\widetilde{C}\lambda ^{\prime
}+i(\chi S^{\ast }+i\sigma ^{\mu }\overline{\chi }A_{\mu })\right] +%
\overline{{\large \chi }}_{\Lambda }[2\widetilde{C}\overline{\lambda
^{\prime }}-i(\overline{\chi }S-i\chi \sigma ^{\mu }A_{\mu })]  \notag \\
&&+\text{ }\frac{1}{2}{\large S}_{\Lambda }\left( \widetilde{C}S^{\ast }+%
\frac{i}{2}\overline{\chi }\overline{\chi }\right) +\frac{1}{2}{\large S}%
_{\Lambda }^{\ast }\left( \widetilde{C}S-\frac{i}{2}\chi \chi \right)  \notag
\\
&&+\text{ }2{\large A}_{\Lambda }^{\mu }(\widetilde{C}A_{\mu }-\chi \sigma
_{\mu }\overline{\chi })+2{\large \lambda }_{\Lambda }^{\prime }(\widetilde{C%
}\chi )+2\overline{{\large \lambda }}_{\Lambda }^{\prime }(\widetilde{C}%
\overline{\chi })+\frac{1}{2}{\large D}_{\Lambda }^{\prime }\widetilde{C}^{2}
\label{lm1}
\end{eqnarray}%
where
\begin{equation}
{\large C}_{\Lambda },\text{ }{\large \chi }_{\Lambda },\text{ }{\large S}%
_{\Lambda },\text{ }{\large A}_{\Lambda }^{\mu },\text{ }{\large \lambda }%
_{\Lambda }^{\prime }={\large \lambda }_{\Lambda }+\frac{i}{2}\sigma ^{\mu
}\partial _{\mu }\overline{{\large \chi }}_{\Lambda },\text{ }{\large D}%
_{\Lambda }^{\prime }={\large D}_{\Lambda }+\frac{1}{2}\partial ^{2}{\large C%
}_{\Lambda }  \label{comp}
\end{equation}%
are the component fields of the Lagrange multiplier superfield $\Lambda
(x,\theta ,\overline{\theta })$ in the standard parametrization (\ref{par}).
Varying the Lagrangian (\ref{ext}) with respect to these fields and properly
combining their equations of motion
\begin{equation}
\frac{\partial \mathcal{L}_{tot}}{\partial \left( {\large C}_{\Lambda },%
{\large \chi }_{\Lambda },{\large S}_{\Lambda },{\large A}_{\Lambda }^{\mu },%
{\large \lambda }_{\Lambda },{\large D}_{\Lambda }\right) }=0
\end{equation}%
we find the constraints which put on the $V$ superfield components
\begin{equation}
C=C_{0},\text{ \ }\chi =0,\text{\ \ }A_{\mu }A^{\mu }=SS^{\ast }\text{,}
\label{const1}
\end{equation}%
being solely determined by the spontaneous SUSY breaking in the visible
sector (\ref{der})
\begin{equation}
P_{C}^{\prime }|_{C=C_{0}}=0\text{ .}  \label{der1}
\end{equation}%
Again, as before in non-SUSY case (\ref{lag}), we only take a solution with
initial values for all fields (and their momenta) chosen so as to restrict
the phase space to vanishing values of the multiplier component fields (\ref%
{comp}) that will provide a ghost-free theory with a positive Hamiltonian.

Now substituting the constraints (\ref{const1}, \ref{der1}) into the total
Lagrangian $\mathcal{L}_{\mathfrak{tot}}$ (\ref{ext}, \ref{lag3}) we
eventually come to the basic Lagrangian in the broken SUSY phase%
\begin{equation}
\mathcal{L}_{tot}^{^{br}}=-\text{ }\frac{1}{4}F^{\mu \nu }F_{\mu \nu
}+i\lambda \sigma ^{\mu }\partial _{\mu }\overline{\lambda }+\frac{1}{2}%
D^{2}+P(C_{0})D\text{ , \ }A_{\mu }A^{\mu }=SS^{\ast }  \label{fin}
\end{equation}%
being supplemented by by the vector field constraint, as indicated. So, for
the constrained vector superfield,
\begin{equation}
\widehat{V}(x,\theta ,\overline{\theta })=C_{0}+\frac{i}{2}\theta \theta S-%
\frac{i}{2}\overline{\theta }\overline{\theta }S^{\ast }-\theta \sigma ^{\mu
}\overline{\theta }A_{\mu }+i\theta \theta \overline{\theta }\overline{%
\lambda }-i\overline{\theta }\overline{\theta }\theta \lambda +\frac{1}{2}%
\theta \theta \overline{\theta }\overline{\theta }D,  \label{sup}
\end{equation}%
we have the almost standard SUSY QED Lagrangian with the same states -
photon, photino and an auxiliary scalar $D$ field - in its gauge
supermultiplet, while another auxiliary complex scalar field $S$ gets only
involved in the vector field constraint. The linear (Fayet-Iliopoulos) $D$%
-term with the effective coupling constant $P(C_{0})$ in (\ref{fin}) shows
that the supersymmetry in the theory is spontaneosly broken due to which the
$D$ field acquires the vev, $D=-P(C_{0})$. Taking the nondynamical $S$ field
in the constraint (\ref{const1}) to be some constant background field (for a
more formal discussion, see below) we come to the SLIV\ constraint (\ref%
{const}) which we discussed above regarding an ordinary non-supersymmetric
QED theory (sec.1). As is seen from this constraint in (\ref{fin}), one may
only have a time-like SLIV in the SUSY framework but never a space-like one.
There also may be a light-like SLIV, if the $S$ field vanishes\footnote{%
Indeed, this case, first mentioned in \cite{nambu}, may also mean
spontaneous Lorentz violation with a nonzero vev $<A_{\mu }>$ $=(\widetilde{M%
},0,0,\widetilde{M})$ and Goldstone modes $A_{1,2}$ and $(A_{0}+A_{3})/2$\ $-%
\widetilde{M}.$ The "effective" Higgs mode $(A_{0}-A_{3})/2$ can be then
expressed through Goldstone modes so that the light-like condition $A_{\mu
}^{2}=0$ is satisfied.}. So, any possible choice for the $S$ field
corresponds to the particular gauge choice for the vector field $A_{\mu }$
in an otherwise gauge invariant theory. Thus, a massless photon emerging
first as a companion of a massless photino (being Goldstone fermion in the
broken SUSY phase) remains massless due to this gauge invariance.

We conclude by showing that our extended Lagrangian $\mathcal{L}_{tot}$ (\ref%
{ext}, \ref{lag3}), underlying the emergent QED model, is SUSY invariant,
and also the constraints (\ref{const1})\ on the field space appearing due to
the Lagrange multiplier term in (\ref{ext}) are consistent with the
supersymmetry. The first part of this assertion is somewhat immediate since
the Lagrangian $\mathcal{L}_{tot}$, aside from the standard supersymmetric
QED part $L_{SQED}$ (\ref{slag}), only contains $D$-terms of various vector
superfield products. They are, by definition, invariant under conventional
SUSY transformations \cite{wess} which for the component fields (\ref{par})
of a general superfield $V(x,\theta ,\overline{\theta })$ (\ref{par}) are
witten as
\begin{eqnarray}
\delta _{\xi }C &=&i\xi \chi -i\overline{\xi }\overline{\chi }\text{ , \ }%
\delta _{\xi }\chi =\xi S+\sigma ^{\mu }\overline{\xi }(\partial _{\mu
}C+iA_{\mu })\text{ , \ }\frac{1}{2}\delta _{\xi }S=\overline{\xi }\overline{%
\lambda }+\overline{\sigma }_{\mu }\partial ^{\mu }\chi \text{ ,}  \notag \\
\delta _{\xi }A_{\mu } &=&\xi \partial _{\mu }\chi +\overline{\xi }\partial
_{\mu }\overline{\chi }+i\xi \sigma _{\mu }\overline{\lambda }-i\lambda
\sigma _{\mu }\overline{\xi }\text{ , \ }\delta _{\xi }\lambda =\frac{1}{2}%
\xi \sigma ^{\mu }\overline{\sigma }^{\nu }F_{\mu \nu }+\xi D\text{ ,}
\notag \\
\delta _{\xi }D &=&-\xi \sigma ^{\mu }\partial _{\mu }\overline{\lambda }+%
\overline{\xi }\sigma ^{\mu }\partial _{\mu }\lambda \text{ .}  \label{trans}
\end{eqnarray}%
However, there may still be left a question whether the supersymmetry
remains in force when the constraints (\ref{const1})\ on the field space are
"switched on" thus leading to the final Lagrangian $\mathcal{L}%
_{tot}^{^{br}} $ (\ref{fin}) in the broken SUSY phase with the both
dynamical fields $C$ and $\chi $ eliminated. This Lagrangian appears similar
to the standard supersymmetric QED taken in the Wess-Zumino gauge, except
that the supersymmetry is spontaneously broken in our case. In the both
cases the photon stress tensor $F_{\mu \nu }$, photino $\lambda $ and
nondynamical scalar $D$ field form an irreducible representation of the
supersymmetry algebra (the last two line in (\ref{trans})). Nevertheless,
any reduction of component fields in the vector superfield is not consistent
in general with the linear superspace version of supersymmetry
transformations, whether it be the Wess-Zumino gauge case or our constrained
superfield (\ref{sup}). Indeed, a general SUSY transformation does not
preserve the Wess-Zumino gauge: a vector superfield in this gauge acquires
some extra terms when being SUSY transformed. The same occurs with our
constrained superfield as well. The point, however, is that in the both
cases a total supergauge transformation
\begin{equation}
V\rightarrow V+i(\Omega -\Omega ^{\ast })\text{ ,}  \label{spg}
\end{equation}%
where $\Omega $ is a chiral superfield gauge transformation parameter, can
always restore the superfield initial form. Actually, the only difference
between these two cases is that whereas the Wess-Zumino supergauge leaves an
ordinary gauge freedom untouched, in our case this gauge is unambiguously
fixed in terms of the above \ vector field constraint (\ref{const1}).
However, this constraint is valid under SUSY transformations provided that
the scalar field components $\varphi $ and $F$ in the $\Omega $ are properly
chosen. Actually, the non-trivial part of the $\widehat{V}$ superfield
transformation which can not be gauged away from the emergent theory (\ref%
{fin}) has the form%
\begin{equation}
\widehat{V}\rightarrow \widehat{V}+i\theta \theta F-i\overline{\theta }%
\overline{\theta }F^{\ast }-2\theta \sigma ^{\mu }\overline{\theta }\partial
_{\mu }\varphi \text{ .}  \label{sg}
\end{equation}%
according to which its vector and scalar field components transform as
\begin{equation}
A_{\mu }\rightarrow A_{\mu }^{\prime }=A_{\mu }-\partial _{\mu }(2\varphi )%
\text{ , \ \ }S\rightarrow S^{\prime }=S+2F\text{ .}  \label{gra}
\end{equation}%
It can be immediately seen that our basic Lagrangian $\mathcal{L}%
_{tot}^{^{br}}$ (\ref{fin}) being gauge invariant and containing no the
scalar $S$ field is automatically invariant under either of these two
transformations individually. In contrast, the supplementary vector field
constraint (\ref{const1}), though it is also turned out to be invariant
under supergauge transformations (\ref{gra}), but only if they are made
jointly. Indeed, for any choice of the scalar $\varphi $ in (\ref{gra})
there can always be found such a scalar $F$ (and vice versa) that the
constraint remains invariant%
\begin{equation}
A_{\mu }A^{\mu }=SS^{\ast }\rightarrow A_{\mu }^{\prime }A^{\prime \mu
}=S^{\prime }S^{\prime \ast }  \label{sc1}
\end{equation}%
In other words, the vector field constraint is invariant under supergauge
transformations (\ref{gra}) but not invariant under an ordinary gauge
transformation. As a result, in contrast to the Wess-Zumino case, the
supergauge fixing in our case will also lead to the ordinary gauge fixing.
We will use this supergauge freedom to reduce the $S$ field to some constant
background value and find the final equation for the gauge function $\varphi
(x)$. So, for the parameter field $F$ chosen in such a way to have
\begin{equation}
S^{\prime }=S+2F=Me^{i\alpha (x)}\text{ },
\end{equation}%
where $M$ is some constant mass parameter (and $\alpha (x)$ is an arbitrary
phase), we come in (\ref{sc1}) to
\begin{equation}
(A_{\mu }-2\partial _{\mu }\varphi )(A^{\mu }-2\partial ^{\mu }\varphi
)=M^{2}\text{ .}
\end{equation}%
that is precisely our old SLIV constraint (\ref{const}) being varied by the
gauge transformation (\ref{gra}). Recall that this constraint, as was
thoroughly discussed in Introduction (sec.1), only fixes gauge (to which
such a gauge function $\varphi (x)$ has to satisfy), rather than physically
breaks gauge invariance.

To summarize, it was shown that the spontaneous SUSY breaking constraints on
the allowed configurations of the physical fields\ (\ref{const1}) in a
general polynomially extended Lagrangian (\ref{ext}) are entirely consistent
with the supersymmetry. In the broken SUSY phase one eventually comes to the
standard SUSY QED type Lagrangian (\ref{fin}) being supplemented by the
vector field constraint invariant under supergauge transformations. One
might think that, unlike the gauge invariant linear (Fayet-Iliopoulos)
superfield term, the quadratic and higher order superfield terms in the
starting Lagrangian (\ref{ext}) would seem to break gauge invariance.
However, this fear proved groundless. Actually, as was shown above in the
section, this breaking amounts to the gauge fixing determined by the
nonlinear vector field constraint mentioned above. It is worth noting that
this constraint formally follows from the SUSY invariant Lagrange multiplier
term in (\ref{ext}) for which is required the phase space to be restricted
to vanishing values of all the multiplier component fields (\ref{comp}). The
total vanishing of the multiplier superfield provides the SUSY invariance of
such restrictions. Any non-zero multiplier component field left in the
Lagrangian would immediately break supersymmetry and, even worse, would
eventually lead to ghost modes in the theory and a Hamiltonian unbounded
from below.

\section{ Spontaneous SUSY breaking in visible and hidden \\ sectors: photino
as pseudo-goldstino}

Let us now turn to matter superfields which have not yet been included in
the model. In their presence the spontaneous SUSY breaking in the visible
sector, which fundamentally underlies our approach, might be
phenomenologically ruled out by the well-known supertrace sum rule \cite%
{wess} for actual masses of quarks and leptons and their superpartners%
\footnote{%
Note that an inclusion of direct soft mass terms for scalar superpartners in
the model would mean in general that the visible SUSY sector is explicitly,
rather than spontaneosly, broken that would immediately invalidate the whole
idea of the massless photons as the zero Lorentzian modes triggered by the
spontaneously broken supersymmetry.}. However, this sum rule is acceptably
relaxed when taking into account large radiative corrections to masses of
supersymmetric particles that proposedly stem from the hidden sector. This
is just what one may expect in conventional supersymmetric theories with the
standard two-sector paradigm, according to which a hidden sector is largely
responsible for SUSY breaking, and the visible sector feels this SUSY
breaking indirectly via messenger fields \cite{wess}. In this way SUSY can
indeed be spontaneously broken at the tree level as well that ultimately
leads to a double spontaneous SUSY breaking pattern in the model considered.

We may suppose, just for uniformity, only $D$-term SUSY breaking both in
visible and hidden sectors\footnote{%
In general, both $D$- and $F$-type terms can be simultaneously used in the
visible and hidden sectors\ (usually just $F$-term SUSY breaking is used in
both sectors \cite{wess}).}. Properly, our supersymmetric QED model may be
further extended by some extra local $U^{\prime }(1)$ symmetry which is
proposed to be broken at very high energy scale $M^{\prime }$ (for some
appropriate anomaly mediated scenario, see \cite{ross} and references
therein). It is natural to think that due to the decoupling theorem all
effects of the $U^{\prime }(1)$ are suppressed at energies \ $E<<M^{\prime }$
by powers of $1/M^{\prime }$ and only the $D^{\prime }$-term of the
corresponding vector superfield $V^{\prime }(x,\theta ,\overline{\theta })$
remains in essence when going down to low energies. Actually, this term with
a proper choice of messenger fields and their couplings naturally provides
the $M_{SUSY\text{ \ }}$order contributions to masses of scalar
superpartners.

As a result, the simplified picture discussed above (in sections 2 and 3) is
properly changed: a strictly massless fermion eigenstate, the true goldstino
$\zeta _{g}$, should now be some mix of the visible sector photino $\lambda $
and the hidden sector\ goldstino $\lambda ^{\prime }$%
\begin{equation}
\zeta _{g}=\frac{\left\langle D\right\rangle \lambda +\left\langle D^{\prime
}\right\rangle \lambda ^{\prime }}{\sqrt{\left\langle D\right\rangle
^{2}+\left\langle D^{\prime }\right\rangle ^{2}}}\text{ .}  \label{tr}
\end{equation}%
where $\left\langle D\right\rangle $ and $\left\langle D^{\prime
}\right\rangle $ are the corresponding $D$-component vevs in the visible and
hidden sectors, respectively. Another orthogonal combination of them may be
referred to as the pseudo-goldstino $\zeta _{pg}$,
\begin{equation}
\zeta _{pg}=\frac{\left\langle D^{\prime }\right\rangle \lambda
-\left\langle D\right\rangle \lambda ^{\prime }}{\sqrt{\left\langle
D\right\rangle ^{2}+\left\langle D^{\prime }\right\rangle ^{2}}}\text{ .}
\label{ps}
\end{equation}%
In the supergravity context, the true goldstino $\zeta _{g}$ is eaten
through the super-Higgs mechanism to form the longitudinal component of the
gravitino, while the pseudo-goldstino $\zeta _{pg}$ gets some mass
proportional to the gravitino mass from supergravity effects. Due to large
soft masses required to be mediated, one may generally expect that SUSY is
much stronger broken in the hidden sector than in the visible one, $%
\left\langle D^{\prime }\right\rangle >>$ $\left\langle D\right\rangle $,
that means in turn the pseudo-goldstino $\zeta _{pg}$ is largely the photino
$\lambda ,$
\begin{equation}
\zeta _{pg}\simeq \lambda \text{ .}
\end{equation}%
These pseudo-goldstonic photinos seem to be of special observational
interest in the model that, apart from some indication of the QED emergence
nature, may shed light on SUSY breaking physics. The possibility that the
supersymmetric Standard Model visible sector might also spontaneously break
SUSY thus giving rise to some pseudo-goldstino state was also considered,
though in a different context, in \cite{vis,tha}. Normally, if the visible
sector possesses the $R$-symmetry which is preserved in the course of the
mediation, then the pseudo-goldstino mass is protected up to the
supergravity effects which violate $R$-symmetry. As a result, the
pseudo-goldstino mass appears proportional to the gravitino mass, and,
eventually, the same region of parameter space simultaneously solves both
gravitino and pseudo-goldstino overproduction problems in the early universe
\cite{tha}.

Apart from cosmological problems, many other sides of new physics related to
pseudo-goldstinos appearing through the multiple SUSY breaking were also
studied recently (see \cite{vis,tha,gol} and references therein). The point,
however, is that there have been exclusively used non-vanishing $F$-terms as
the only mechanism of the visible SUSY breaking in models considered. In
this connection, our pseudo-goldstonic photinos solely caused by
non-vanishing $D$-terms in the visible SUSY sector may lead to somewhat
different observational consequences. One of the most serious differences
belongs to Higgs boson decays provided that our QED model is further
extended to supersymmetric Standard Model. For the cosmologically safe
masses of pseudo-goldstino and gravitino ($\lesssim $ $1keV$, as typically
follows from $R$-symmetric gauge mediation) these decays are appreciably
modified. Actually, the dominant channel becomes the conversion of the Higgs
boson (say, the lighter CP-even Higgs boson $h^{0}$) into a conjugated pair
of corresponding pseudo-sgoldstinos $\phi _{pg}$ and $\overline{\phi }_{pg}$
(being superpartners of\ pseudo-goldstinos $\zeta _{pg}$ and $\overline{%
\zeta }_{pg}$, respectively), $h^{0}\rightarrow \phi _{pg}+\overline{\phi }%
_{pg}$, once it is kinematically allowed. This means that the Higgs boson
will dominantly decay invisibly for $F$-term SUSY breaking in a visible
sector \cite{tha}. By contrast, for the $D$-term SUSY breaking case
considered here the roles of pseudo-goldstino and pseudo-sgoldstino are just
played by photino and photon, respectively, that could make the standard
two-photon decay channel of the Higgs boson to be even somewhat enhanced. In
the light of recent discovery of the Higgs-like state \cite{h} just through
its visible decay modes, the $F$-term SUSY breaking in the visible sector
seems to be disfavored by data, while $D$-term SUSY breaking is not in
trouble with them. \

\section{Concluding remarks}

It is well known that spontaneous Lorentz violation in general vector field
theories may lead to an appearance of massless Nambu-Goldstone modes which
are identified with photons and other gauge fields in the Standard Model.
Nonetheless, it may turn out that SLIV is not the only reason for emergent
massless photons to appear, if spacetime symmetry is further enlarged. In
this connection, a special link may be related to supersymmetry that we
tried to argue here by the example of supersymmetric QED that can be then
straightforwardly extended to the Standard Model.

The main conclusion which has appeared in the SUSY context is that
spontaneous Lorentz violation caused by an arbitrary potential of vector
superfield $V(x,\theta ,\overline{\theta })$ never goes any further than
some noncovariant gauge constraint put on its vector field component $A_{\mu
}(x)$\ associated with a photon. This allows to think that physical Lorentz
invariance is somewhat protected by SUSY, thus only admitting the
"condensation" of the gauge degree of freedom in the vector field $A_{\mu }$%
. The point, however, is that even in this case when SLIV is "inactive" it
inevitably leads to the generation of massless photons as vector
Nambu-Goldstone modes provided that SUSY itself is spontaneously broken. In
this sense, a generic trigger for massless photons to dynamically emerge
happens to be spontaneously broken supersymmetry rather than physically
manifested Lorentz noninvariance.

To see how this idea may work we considered supersymmetric QED model
extended by an arbitrary polynomial potential of a general vector superfield
that induces the spontaneous SUSY violation in the visible sector. In the
broken SUSY phase one eventually comes to the standard SUSY QED type
Lagrangian (\ref{fin}) being supplemented by the vector field constraint
invariant under supergauge transformations. As result, a massless photon
appears as a companion of a massless photino which emerges in fact as the
Goldstone fermion state in the tree approximation. However, being mixed with
another goldstino appearing from a spontaneous SUSY violation in the hidden
sector this state largely turns into the light pseudo-goldstino. Remarkably,
the photon masslessness appearing at the tree level is further protected
against radiative corrections by the simultaneously generated special gauge
invariance. This invariance is only restricted by the nonlinear gauge
condition put on vector field values, $A_{\mu }A^{\mu }=|S|^{2}$, so that
any possible choice for the nondynamical $S$ field corresponds to the
particular gauge choice for the vector field $A_{\mu }$ in an otherwise
gauge invariant theory. The point, however, is that this nonlinear gauge
condition happens at the same time to be the SLIV type constraint which
treats in turn the physical photon as the Lorentzian NG mode. So,
figuratively speaking, the photon passes through three evolution stages
being initially the massive vector field component of a general vector
superfield (\ref{lag3}), then the three-level massless companion of the
Goldstonic photino in the broken SUSY stage (\ref{der}) and finally the
generically massless state as the emergent Lorentzian mode in the inactive
SLIV stage (\ref{const1}).

As to pseudo-goldstonic photinos appeared in the model, they seem to be of
special observational interest that, apart from some indication of the QED
emergence nature, may appreciably extend the scope of SUSY breaking physics
being actively discussed in recent years. In contrast to all previous
considerations with non-vanishing $F$-terms as a mechanism of visible SUSY
breaking, our pseudo-goldstonic photinos caused by non-vanishing $D$-terms
in the visible SUSY sector will lead to somewhat different observational
consequences. These and related points certainly deserve to be explored in
greater detail.

\section*{Acknowledgements}

I thank Colin Froggatt, Alan Kostelecky, Rabi Mohapatra and Holger Nielsen
for stimulating discussions and correspondence. Discussions with the
participants of the Workshop "What Comes Beyond the Standard Models?"
(14--21 July 2013, Bled, Slovenia), especially with Eduardo Guendelman and
Norma Mankoc, were also extremely interesting and useful.

\end{document}